\documentclass[12pt,preprint]{aastex}

\begin{document}

\title{On the nature of radio pulsars with long periods}
\shorttitle{On the nature of long period radio pulsars}
\shortauthors{Lomiashvili et al.}
\author{D. Lomiashvili}
\affil{Tbilisi state university, Chavchavadze avenue 3, 0128,
Tbilisi, Georgia} \email{lomso@mail.ru}
\author{G. Machabeli}
\affil{Abastumani Astrophysical Observatory, Al. Kazbegi Avenue 2a,
0160, Tbilisi, Georgia} \email{g\_machabeli@hotmail.com}
\author{I. Malov}
\affil{Pushchino Radio Astronomy Observatory, P.N.Lebedev
Institute of Physics, 142290, Pushchino, Moscow Region, Russia,
and Isaac Newton Institute of Chile, Pushchino Branch}
\email{malov@prao.psn.ru}
\begin{abstract} It is shown that the drift
waves near the light cylinder can cause the modulation of emission
with periods of order several seconds. These periods explain the
intervals between successive pulses observed in ''magnetars'' and
radio pulsars with long periods. The model under consideration gives
the possibility to calculate real rotation periods of host neutron
stars. They are less than 1 sec for the investigated objects. The
magnetic fields at the surface of the neutron star are of order
10$^{11}$ - 10$^{13}$ G and equal to the fields usual for known
radio pulsars.
\end{abstract}
\section{Introduction}

According to the plasma model pulsar radio emission is generated
in the
electron-positron plasma, which appears by the avalanche process of (e$^{+}$e%
$^{-}$) pair production. The accomplishment of this process
requires fulfilment of the following condition:
\begin{equation}
\hspace{6pt} B_{Schw} \geq B_{s}\geq B_{dl}
\end{equation}
Here $B_{s}=3.2\times 10^{19}\sqrt{\dot{P}P}$ is pulsar surface
magnetic field inferred from observational values of spin period
and period derivative, $B_{Schw}=m_{e}^{2}c^{3}/e\hbar \approx
4.4\times 10^{13}G$ is Schwinger limit i.e. the magnetic field at
which the electron cyclotron energy equals the electron rest mass
energy and $B_{dl}$ is the magnetic field corresponding to the
so-called "death line" (see Fig. 1) \citep{young99}. The various
configurations of the surface magnetic field correspond to
different death lines. They depend not only on magnetic field
configuration, but significantly on whether the origin of gamma
quanta, which are responsible for pair production, is curvature
radiation or inverse Compton scattering \citep{zhang00}. As a
mechanism of producing of gamma quanta we took curvature radiation
with sunspot configuration of magnetic field. In this case "death
line" is defined by the condition that the potential drop across
the gap required to produce enough pairs per primary to screen out
the parallel electric field is larger than the maximum potential
drop available from the pulsar. The quantity $B_{dl}$ can be
solved from the following equation \citep{chen93}:
\begin{equation}
7\log B_{dl}-13\log P=78
\end{equation}

Recently there were discovered three long period radio pulsars PSR
J2144-3933 \citep{young99}, PSR J1847-0130 \citep{mcl03} and PSR
J1814-1744 \citep{camilo00}(see Table 1), that break the condition
above (the first one breaks the right-hand side of the inequality
and the other two break the left). PSR J2144-3933 is distinguished
by some other characteristics. It has the lowest spindown
luminosity ($\dot{E}=4\pi^{2}I\dot{P}/P^{3}\approx 3.2\times
10^{28}erg\cdot s^{-1}$) of any known pulsar. The beaming fraction
(that is, the fraction of the celestial sphere swept across by the
beam) is also the smallest, $W\approx 1/300$. On the other hand,
PSR J1847-0130 \citep{mcl03} and PSR J1814-1744 \citep{camilo00}
are isolated radio pulsars having the largest, "magnetar"-like,
inferred surface dipole magnetic fields yet seen in the
population: $9.4\times 10^{13}$ G and $5.5\times 10^{13}$ G,
respectively. These pulsars show apparently normal radio emission
in the regime of magnetic field strength ($B_{s}\geq B_{Schw}$),
where plasma models predict no emission should occur. However, the
nature of the Schwinger limit is not clear and is the subject of
long term discussions. \citet{bar01} proposed that for extremely
strong fields, photon splitting dominates over pair production and
the corresponding death line is a function of both field strength
and period. This is definitely neither a certain nor a hard limit.
Furthermore, \citet{usov02} has argued that since only one photon
polarization mode splits, the onset of photon splitting does not
suppress the production of pairs. It is very important to
investigate this interesting problem, but it is beyond the
framework of our model, especially since we argue that magnetic
fields of neutron stars in our investigation are less than the
Schwinger limit.

 A model that explains the phenomenon of radio
emission from these pulsars and all their anomalous properties
does not exist.

An important feature of our model, which provides a natural
explanation of most of the properties of these pulsars, is the
presence of very low frequency, nearly transverse drift waves
propagating across the magnetic field and encircling the open
field line region of the pulsar magnetosphere
\citep{kaz91b,kaz96}. These waves only periodically change the
direction of the radio emission and are not  directly observable.
We give a description of this model in Sections 2, 3 and 4. Some
estimates of the rotational and angular parameters of these
pulsars are given in Section 5. We discuss the obtained results in
Sections 6 and 7.

\placetable{1}

\section{Emission model}

As mentioned the pulsar magnetosphere is filled by a dense
relativistic electron-positron plasma. The (e$^{+}$e$^{-}$) pairs
are generated as a consequence of the avalanche process [first
described by \citet{sturrock71}], and they flow along the open
magnetic field lines. The plasma is multicomponent, with a
one-dimensional distribution function
\citep[see][Fig.1]{arons81a}, containing the following:

(i) electrons and positrons of the bulk of plasma with a mean
Lorentz factor of $\gamma _{p}$ and density $n_{p}$;

(ii) particles of the high-energy 'tail' of the distribution function with $%
\gamma _{t}$ and $n_{t}$, stretched in the direction of positive
momentum;

(iii) the ultrarelativistic ($\gamma _{b}\sim 10^{6}$) beam of
primary particles with so-called 'Goldreich-Julian' density
\citep{gold69} $n_{b}\approx 7\times
10^{-2}B_{s}P^{-1}(R_{0}/R)^{3}\left[ cm^{-3}\right] $ (where P is
the pulsar period, $R_{0}\approx 10^{6}$cm is the neutron star
radius, $B_{s}$ is the magnetic field value at the stellar surface
and $R$ is the distance from the center of the neutron star). This
density is much less than the density of secondary pairs $n_{p}$.

 Let us note that such parameters are only possible under the
assumption that strong and curved multipolar fields persist in the
pair-production region, near the stellar surface
\citep{arons81b,mach89}.

 Such a distribution function should generate various wave-modes
under certain conditions. The waves then propagate in the pair
plasma of the pulsar magnetosphere, transform into vacuum
electromagnetic waves as the plasma density drops, enter the
interstellar medium and reach an observer as pulsar radio
emission. This waves leave the magnetosphere propagating at
relatively small angles to the pulsar magnetic field
\citep{kaz91a}.

 Particles moving along the
curved magnetic field undergo drift motion transversely to the
plane where the field line lies. The drift velocity can be written
as
\begin{equation}
u=c\gamma \upsilon_{\varphi}/\omega_{B}R_{c}
\end{equation}
where $\omega_{B}=eB/mc$, $B=B_{s}(R_{0}/R)^{3}$ . $R_{c}$ is the
curvature radius of the dipolar magnetic field line, $\gamma $ is
the relativistic Lorentz factor of a particle and $\upsilon
_{\varphi }$ is the particle velocity along the magnetic field.
Here and below the cylindrical coordinate system ($x,r,\varphi $)
is chosen, with the $x$-axis directed transversely to the plane of
the field line, while $r$ and $\varphi$ are the radial and
azimuthal coordinates, respectively.

 Generation of radio emission is possible if at
least one of the following resonance conditions is fulfilled:
\begin{equation}
\quad\qquad\textrm{cyclotron instability} \qquad \omega
-k_{\varphi }\upsilon _{\varphi }-k_{x}u=\frac{\omega _{B}}{\gamma
_{res}} \qquad
\end{equation}
\begin{equation}
\textrm{Cherenkov instability}\qquad \omega -k_{\varphi}\upsilon
_{\varphi }-k_{x}u=0
\end{equation} These conditions are very sensitive to
the parameters of the magnetospheric plasma, particularly to the
value of the drift velocity (see eq.[3]), and hence to the
curvature of the magnetic field lines \citep{kaz96}.

 It should be noted that, in the absence of drift motion, the ordinary
Cherenkov interaction implies that the phase velocity of a wave
equals the velocity of the particles in both value and direction.
In other words, an observer moving with the same velocity as the
particles, should detect the same phase of the wave for a
sufficiently long time ($\tau \sim 1/\Gamma $ ). This is, however,
impossible for a wave propagating transversely to the particles'
velocity. On the other hand, the drift velocity (equation (3)) is
directed along the phase velocity of such a wave. This allows
wave-particle resonant interaction.

\section{Generation of drift waves}

It has been shown \citep{kaz91b,kaz96} that, in addition to the
waves mentioned above (the characteristic frequencies of which
fall into the radio band) propagating with small angles to the
magnetic field lines, very low frequency, nearly transverse drift
waves can be excited. They propagate across the magnetic field, so
that the angle between ${\bf k}$ and ${\bf B}$
is close to $\pi /2$. In other words, $k_{\bot }/k_{\varphi }\gg 1$, where $%
k_{\bot }=(k_{r}^{2}+k_{\varphi }^{2})^{1/2}$ . Assuming $\gamma
(\omega /\omega _{B})\ll 1$, $(u_{\alpha }/c)^{2}\ll 1$,
$k_{\varphi }/k_{x}\ll 1$ and $k_{r}\rightarrow 0$, we can write
the general dispersion equation of the drift waves in the
following form \citep{kaz91a,kaz91b,kaz96}:
\begin{eqnarray}
\left( 1-\sum\limits_{\alpha }\frac{\omega _{\alpha }^{2}}{\omega
}\int \frac{u_{\alpha }^{2}}{\upsilon _{\varphi }c}\frac{1}{\omega
-k_{\varphi }\upsilon _{\varphi }-k_{x}u_{a}}\frac{\partial
f_{\alpha }}{\partial p}dp- \frac{k_{\varphi }^{2}c^{2}}{\omega
^{2}}\right) \times \nonumber
\\\left( 1+\sum\limits_{\alpha }\frac{\omega
_{\alpha }^{2}}{\omega }\int \frac{ \upsilon _{\varphi }/c}{\omega
-k_{\varphi }\upsilon _{\varphi }-k_{x}u_{a}} \frac{\partial
f_{\alpha }}{\partial p}dp\frac{k_{\varphi }^{2}c^{2}}{\omega
^{2}}\right) - \nonumber
\\
\left( \frac{k_{x}k_{\varphi }c^{2}}{\omega
^{2}}+\sum\limits_{\alpha } \frac{\omega _{\alpha }^{2}}{\omega
}\int \frac{u_{\alpha }/c}{\omega -k_{\varphi }\upsilon _{\varphi
}-k_{x}u_{a}}\frac{\partial f_{\alpha }}{
\partial p}dp\right) ^{2}=0
\end{eqnarray}
where $\alpha $ denotes the sort of particles (electrons or
positrons) and $\omega _{\alpha }^{2}=4\pi n_{\alpha }e^{2}/m$,
$f_{\alpha }$ is the distribution function and $p$ is the momentum
of the plasma particles.

Let us assume that
\begin{equation}
\omega =k_{\varphi }\upsilon _{\varphi }+k_{x}u_{b}+i\Gamma
\end{equation}
where $u_{b}$ is the drift velocity of the beam
particles (see eq.[3]). In the approximation $k_{\varphi }\upsilon _{\varphi }\ll k_{x}u_{b}$ and $%
k_{x}^{2}\ll \omega _{p}^{2}/\gamma _{p}^{3}c^{2}$, the imaginary
part can be written as
\begin{equation}
\Gamma = Im \omega \approx \left( \frac{n_{b}}{n_{p}}\right)
^{1/2}\left( \frac{\gamma _{p}^{3}}{\gamma _{b}}\right)
^{1/2}k_{x}u_{b}
\end{equation}
According to equation (7),
the frequency of a drift wave can be written as
\begin{equation}
\omega _{0}=Re \omega =k_{\varphi }\upsilon _{\varphi
}+k_{x}u_{b}\approx k_{x}u_{b}
\end{equation}
Drift waves propagate across the magnetic field and encircle the
region of the open field lines of the pulsar magnetosphere. They
draw energy from the longitudinal motion of the beam particles, as
in the case of the ordinary
Cherenkov wave-particle interaction. However, they are excited only if $%
k_{x}u_{b}\neq 0$, i.e., in the presence of drift motion of the
beam particles. Note that these low-frequency waves are nearly
transverse, with the electric vector being directed almost along
the local magnetic field. Let us note that although $k_{\varphi
}\upsilon _{\varphi }\ll k_{x}u_{b}$ for the drift waves, there
still exists a nonzero $k_{\varphi }$. It appears that growth rate
(equation (8)) is rather small. However, the drift waves propagate
nearly transversely to the magnetic field, encircling the
magnetosphere, and stay in the resonance region for a substantial
period of time. Although the particles give a small fraction of
their energy to the waves and then leave the interaction region,
they are continuously replaced by the new particles entering this
region. The waves leave the resonance region considerably slower
than the particles. Hence, there is no sufficient time for the
inverse action of the waves on the particles. The accumulation of
energy in the waves occurs without quasi-linear saturation. The
amplitude of the waves grows until the nonlinear processes
redistribute the energy over the spectrum. As was demonstrated by
\citet{kaz91b}, the strongest nonlinear process in this case is
the induced scattering of waves on plasma particles. Therefore,
the growth of the drift-wave amplitude continues until the
decrement of the nonlinear waves $\Gamma _{NL}$ becomes equal to
the linear decrement $\Gamma $. As a result, one obtains
quasi-regular configurations of drift waves. Generally, the
nonlinear scattering pumps the wave energy into the
long-wavelength domain of the spectrum.
\begin{equation}
\lambda _{\max }\approx r_{LC}=\frac{cP}{2\pi }
\end{equation}
Here $r_{LC}$ is the radius of the light cylinder.

According to equations (9), (10) and (3), the period of the drift
waves can be written as:
\begin{equation}
P_{dr}=\frac{2\pi }{\omega _{dr}}=\frac{2\pi }{k_{x}u_{b}}=\frac{\lambda }{%
u_{b}}=\frac{e}{4\pi ^{2}mc}\frac{BP^{2}}{\gamma _{b}}
\end{equation}
It appears that the period of the drift wave is of the order of
several seconds. It is possible to determine the relationship
between $P_{dr}$, the derivative and the rate of slowing down of
the neutron star from equation (11)
\begin{equation}
\dot P_{dr}=\frac{eB}{2\pi ^{2}mc\gamma _{b}}P \dot P
\end{equation}
For the considered values of the parameters we obtain $\dot
P_{dr}\approx10 \dot P$ This relation is kept during the entire
life of the pulsar, until it stops emitting.

\section{Mechanism of field line curvature change}

Let us assume that a drift wave with the dispersion defined by
equation (7) is excited at some place in the pulsar magnetosphere.
It follows from the Maxwell equations that $B_{r}=E_{\varphi
}(k_{x}c/\omega )$, hence $B_{r}\gg E_{\varphi }$ for such a wave.
Therefore, excitation of a drift wave causes particular growth of
the $r$-component of the local magnetic field.

The field line curvature $\rho _{c}\equiv 1/R_{c}$ is defined in a
Cartesian frame of coordinates $(x, y,z)$ (where $z$ axis is
directed perpendicular to the plane of the field line) as
\begin{equation}
\rho _{c}=\left[ 1+\left( \frac{dy}{dx}\right) ^{2}\right] ^{-3/2}\frac{%
d^{2}y}{dx^{2}}
\end{equation}
where $dy/dx=B_{y}/B_{x}$. Using $\left( \nabla \cdot{\bf
B}\right) =0$ and rewriting equation (13) in  cylindrical
coordinates, we obtain
\begin{equation}
\rho _{c}=\frac{1}{r}\frac{B_{\varphi }}{B}-\frac{1}{r}\frac{1}{B}\frac{%
B_{\varphi }^{2}}{B^{2}}\frac{\partial B_{r}}{\partial \varphi }
\end{equation}

Here $B=\left( B_{\varphi }^{2}+B_{r}^{2}\right) ^{1/2}\approx
B_{\varphi }\left[ 1+B_{r}^{2}/2B_{\varphi }^{2}\right] $.
Assuming that $k_{\varphi }r\gg 1$, we obtain from equation (14)
\begin{equation}
\rho _{c}=\frac{1}{r}\left( 1-k_{\varphi }r\frac{B_{r}}{B_{\varphi
}}\right)
\end{equation}
From equation (15) it is clear that even a small change of $B_{r}$
causes a significant change of $\rho _{c}$. The variation of the
field line curvature can be estimated as
\begin{equation}
\frac{\Delta \rho_{c}}{\rho_{c}}\approx k_{\varphi }r\frac{\Delta
B_{r}}{ B_{\varphi }}
\end{equation}
It follows that even a drift wave with a modest amplitude
$B_{r}\sim \Delta B_{r}\sim 0.01B_{\varphi }$ alters the field
line curvature substantially, $\Delta \rho_{c}/\rho_{c}\sim 0.1$

Since radio waves propagates along the local magnetic field lines,
curvature variation causes a change of the emission direction.

\section{The model}

There is unequivocal correspondence between the observable intensity and $%
\alpha $ (the angle between the observer's line of sight and the
emission direction (see Fig. 2)). The maximum of the intensity
corresponds to the minimum of $\alpha $. The period of a pulsar is
the time interval between the neighboring maximums of the
observable intensity (minimums of $\alpha $). According to this
fact, we can say that the observable period is representative of
the value of $\alpha $ and as it appears below, it might differ
from the spin period of the pulsar.

\placefigure{2}
\begin{equation}
\cos \alpha = {\bf A}\cdot{\bf K}
\end{equation}
where ${\bf A }$ and ${\bf K}$ are unit guide vectors of
observer's and emission axes, respectively. In the spherical
coordinate system ($r,\varphi ,\theta $), combined with the plane
of the pulsar rotation, these vectors can be expressed as:
\begin{equation}
{\bf A}=\left( 1,0,\delta \right)
\end{equation}
\begin{equation}
{\bf K}=\left( 1,\Omega t,\beta \right)
\end{equation}
where $\Omega =2\pi/P$ is the angular velocity of the pulsar. $%
\delta $ is the angle between the rotation and observer's axes,
and $\beta $ is the angle between the rotation and emission axes
(see Fig. 2).

From equations (17),(18) and (19), it follows that:
\begin{equation}
\alpha =\arccos \left( \sin \delta \sin \beta \cos \Omega t+\cos
\delta \cos \beta \right)
\end{equation}
In the absence of the drift wave $\beta =\beta _{0}=const$, and
consequently the period of $\alpha $ is equal to $ 2\pi/\Omega $.

According to equation (16), in the presence of the drift wave, the
fractional variation $\Delta \rho_{c}/\rho_{c}$ is proportional to
the magnetic field of the wave $B_{r}$, which is periodically
changing. So $\beta =\beta \left( t\right) $ is harmonically
oscillating about $\beta _{0}$ with an amplitude $\Delta \beta
=\Delta \rho_{c}/\rho_{c}$ and rate $\omega _{dr}=2\pi /P_{dr}$ .
Thus, we can write that
\begin{equation}
\beta =\beta _{0}+\Delta \beta \sin \left( \omega _{dr}t+\varphi
\right)
\end{equation}
According to equations (20) and (21,) we obtain
\begin{equation}
\alpha =\arccos \left( \sin \delta \sin \left( \beta _{0}+\Delta
\beta \sin \left( \omega _{dr}t+\varphi \right) \right) \cos
\Omega t+\cos \delta \cos \left( \beta _{0}+\Delta \beta \sin
\left( \omega _{dr}t+\varphi \right) \right) \right)
\end{equation}
\placefigure{3}
\begin{equation}
\alpha _{\min }^{k}=\left| \left( \beta _{0}-\delta \right)
+\Delta \beta \sin \left( 2\pi k\frac{\omega _{dr}}{\Omega
}+\varphi \right) \right|
\end{equation}
$\alpha_{\min }^{k}$ is the minimum of $\alpha$ after $k$
revolutions of the pulsar \footnote{The detection moment of any
pulse is taken as the zero point of the time reckoning}. The
parameters of the pulse profile (e.g. width, height etc.)
significantly depend on what the minimal angle would be between
the  emission axis and the observer's axis while first one passes
the other (during one revolution). If the emission cone does not
cross the observer's line of
sight entirely (i.e., the minimal angle between them is more than cone angle $%
\vartheta $, see inequality (24a)), then we cannot observe the
pulsar emission. On the other hand, inequality (24b) defines
condition that is necessary for emission detection.
\begin{mathletters}
\begin{equation}
\alpha _{\min }^{k}>\vartheta
\end{equation}
\begin{equation}
\alpha _{\min }^{k}<\vartheta
\end{equation}
\end{mathletters}
Hence, for some values of parameters $\Omega$, $\omega_{dr}$, $\beta $, $%
\Delta \beta $, $\delta $, $\varphi $ and $\vartheta $ (Set A) it
is possible to accomplish the following regime: after every $k=m$
turn, the minimal value of $\alpha $ ($\alpha _{\min }^{m}$ )
satisfies condition (24b) while for intervening values of $k$
($1\leq k\leq (m-1)$, where  $k$ and $m$ are positive integers),
$\alpha _{\min }^{m}$ satisfies condition (24a). In that case the
observable period $P_{obs}$ does not represent the real pulsar
spin period, but is divisible by it.
\begin{equation}
P_{obs}=mP
\end{equation}
Hence,
\begin{equation}
\dot{P}_{obs}=m\dot{P},
\end{equation}
where $P$ is the pulsar spin period.

The dipolar magnetic field strength on the neutron star surface
can be written as:
\begin{equation}
B=3.2\times 10^{19}\sqrt{\dot{P}P}
\end{equation}
From equations (25),(26) and (27), it follows that
\begin{equation}
B=\frac{B_{obs}}{m}
\end{equation}
After inserting equations (28) and (25) in equation (2), we
obtain:
\begin{equation}
7\log B-13\log P=(7\log B_{obs}-13\log P_{obs})+6\log m \geq 78
\end{equation}
Then
\begin{equation}
6\log m \geq 78-7\log B_{obs}+13\log P_{obs}
\end{equation}
It can be verified that there exists value for $m$ that satisfies
equation (30) and the following condition simultaneously
\begin{equation}
B=\frac{B_{obs}}{m} < B_{Schw}
\end{equation}
Thus, it is possible to fulfil the conditions necessary for
$($e$^{+}$e$^{-})$ pair production for some values of $m$.

There is a common problem for pulsars II and III presented in
Table 1 (their surface magnetic field strength exceeds the quantum
critical value $B>B_{Schw}$), whereas for PSR J2144-3933 the
problem appears in a different way. Equation (29) is not
fulfilled. After substituting $B_{obs}$ and $P_{obs}$ from Table 1
in equation (30), we obtain:
\begin{equation}
6\log m \geq 4
\end{equation}
Thus, if equation (32) is accomplished, it is possible for PSR
J2144-3933 to generate radio emission.

For better estimation of $m$ we can use observational data for
beaming fractions. From Figure 2 appears that the pulse width can
be expressed as:
\begin{equation}
W=P\times2\sin\theta/2\pi\sin\delta ,
\end{equation}
after inserting equation (25) in equation (33), we obtain:
\begin{equation}
mW/P_{obs}=\sin\theta/\pi\sin\delta
\end{equation}
As was mentioned above, to accomplish the described regime
(eq.[25]), (the angle between the observer's line of sight and the
emission direction after one revolution from that moment when they
were coincident ($\alpha=0$, Fig. 3)) $\alpha _{\min }^{m}$ must
exceed $\theta$. Since
\begin{equation}
\alpha _{min }^{1}=\Delta\beta(2\pi P/P_{dr}),
\end{equation}
if we assume that $\beta_{0}=\delta$, then we get
$P_{dr}=2P_{obs}=2mP$ and
\begin{equation}
\theta_{max}=\Delta\beta(\sin\pi/m);
\end{equation}
if we substitute this equation in equation (34), we obtain:
\begin{equation}
W/P_{obs}=\Delta\beta(\sin\pi/m)/m\pi\sin\delta
\end{equation}
Here the left-hand side is known from the observations. Equation
(37) gives us the ability to estimate the angular parameters of
pulsars for the given values of $m$.

If we consider these pulsars in the framework of our model, their
parameters (e.g. spin and angular parameters) will get new 'real'
values, shown in Table 2.

According to the obtained results, the considered pulsars are
placed on (P-B) diagram as shown in Fig. 4

\placefigure{4}

Thus, we developed the theoretical model of pulsar emission, in the
framework of which we explained all specific features of pulsars
presented in Table 1.

\section{Discussion}

It should be noted that this model is applicable to the entire
population of pulsars, but the effects caused by drift waves are
different depending on the values of the parameters (Set A). In
the case of large  $\Delta \beta$ ($\Delta \beta>\vartheta$), the
most interesting effect is the lengthening
of the observable period (see eq.[25]) which is accomplished only when $%
P_{dr}$ is divisible by $P$ to  high accuracy. It explains the
lack of such kinds of pulsars.

In the case of small $\Delta \beta$ ($\Delta \beta<\vartheta$),
the observable period does not increase (except for $\left| \beta
_{0}-\delta \right|\approx \Delta \beta$ ), but some other
interesting effects appear, such as drifting subpulses
\citep{kaz91c} and period and period derivative oscillation
phenomenon, which is observed in PSR B1828-11 \citep{stairs00} and
PSR B1642-03 \citep{shaba01}. Some authors
\citep{jones01,link01,rez03} have proposed different models to
explain this phenomenon within the framework of free precession of
the  neutron star. As shown by \citet{shaham77} and
\citet{sedrak99} the existence of precession in the neutron star
is in strong conflict with the superfluid models for the neutron
star interior structure. Therefore, we can declare that there does
not yet exist a self-consistent explanation of this fact. We plan
to study  this problem in detail in a forthcoming paper.

If $P_{dr}$ is not divisible by $P$, then the observed intensity
must be modulated with the period of the drift wave. It is
impossible to get such variations with integrated pulse profiles.
Deviations of integrable pulse intensities damp each other. The
only possible way to prove this scenario is by single pulse
observations. Such observations really show intensity variations
\citep{kar01}. Although they do not have have a harmonic nature
(this is due to various noises and insufficient resolution), it
benefits our model. So if experiments are modified to evolve the
oscillating component, it would help validate our theory.

Let us consider pulsars with very short periods. As mentioned
above, drift waves arise in the vicinity of the light cylinder.
The shorter the pulsar spin period, the smaller is the radius of
the light cylinder, and consequently. the larger is the magnetic
field value in the wave generation region ($B_{\varphi} \approx
B=B_{s}(R_{0}/R)^{3}$). Thus, if we take this consideration into
account, from equation (16) it follows that
 for  pulsars whose period is much less than $0.1s$, the amplitude
 of the
oscillation in the emission direction would be so small ($\Delta
\beta < 1^{o}$), that the presence of drift waves would not cause
any significant effect.

We believe that if the scenario defined above (enlarging of the
observable period due to emission direction changing by drift
waves) is accomplished then there must be simultaneous modulation
of the radio and X-ray emission. Some evidence of this fact is
detected radio emission from SGR 1900+14 \citep{shitov99} and AXP
1E2259+586 \citep{malof01}. Higher radio frequency emission from
these magnetars has not been observed yet, but discussion about
this issue differs from objectives of our paper. However,
detection of both types of emission is rare event, because they
are generated on different altitudes, radio and X-ray emission
propagate in different directions. This implies that the pulsars
considered in this paper do not show X-ray emission.

\section{Conclusions}
After these considerations we can divide radio pulsars into the
following groups, which are listed with their main requirements:

1. Rapidly rotating pulsars, for which $\Delta\beta$ is too small.
None of the mentioned effects there should exist for these.

2. Pulsars with $\Delta \beta<\vartheta$ and
$(P_{dr}-P)/P_{dr}<<1$. In this case, period, period derivative
and pulse shape oscillation should appear. In the case of low
accuracy of equality between $P_{dr}$ and $P$, subpulse drift can
be observed \citep{kaz91a, gog05}.

3. Pulsars with $\Delta \beta<\vartheta$. They should show
observed intensity variations, modulated with the period of the
drift waves.

4. Pulsars with $\Delta \beta>\vartheta$ and ${(P_{dr}-mP)/P}<<1$
(where $m$ is a positive integer). These appear different from the
real, long observable rotation period.

Thus, long-period radio pulsars represent one of the branches of
usual pulsars and must be considered in the frameworks of
traditional theories for the specific values of the parameters
(Set A).

\acknowledgments We thank the referee for helpful suggestions.
This work was partly supported by grants from the Georgian Academy
of Science, the Russian Foundation for Basic Research (project
03-02-16509), and the NSF (project 00-98685).

\clearpage

\begin{figure}
\epsscale{1} \plotone{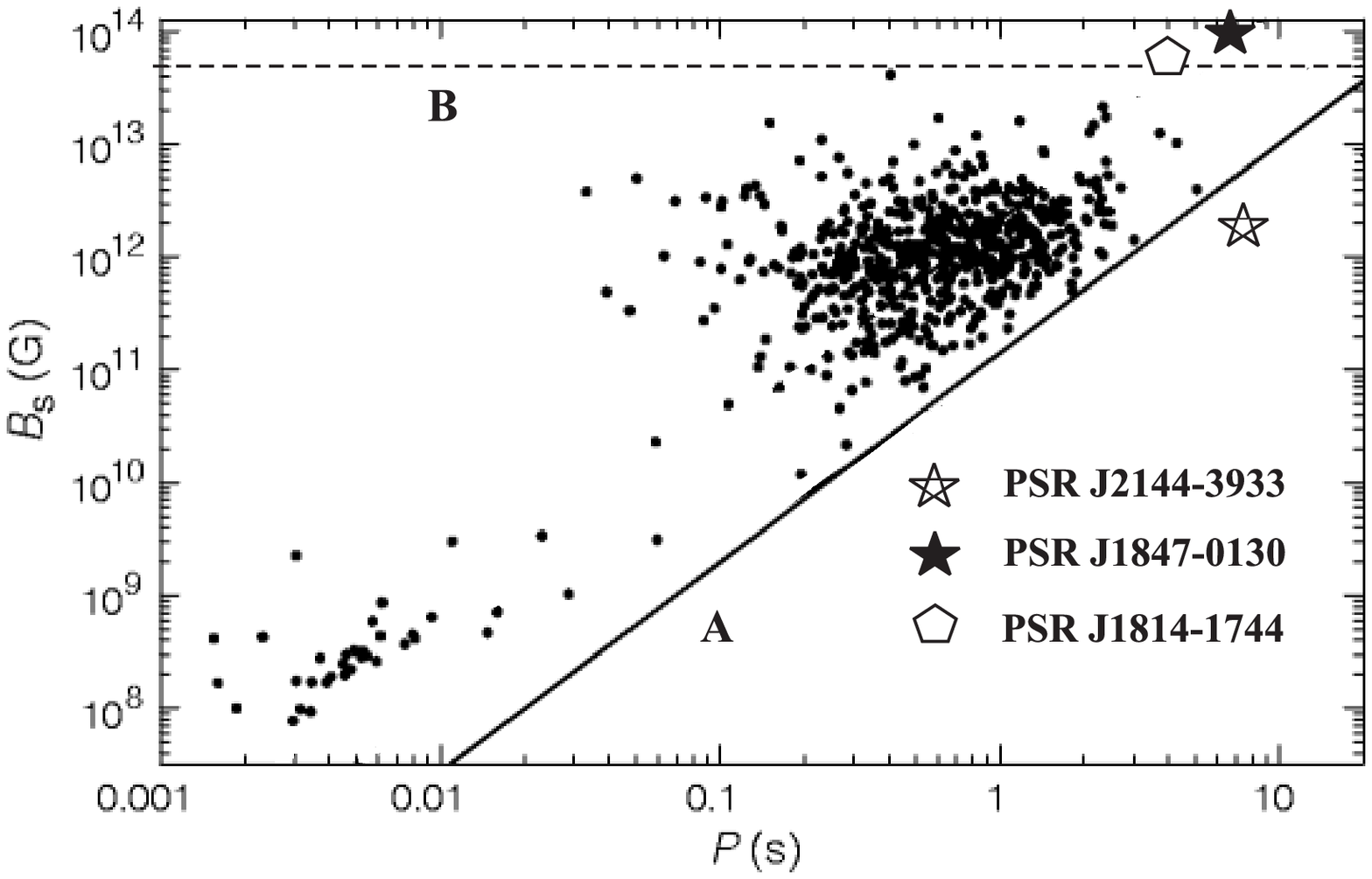} \caption{Line {\bf A}. "Death line"
when curvature radiation with the sunspot configuration of the
magnetic field is taken as the mechanism of producing the gamma
quanta. Line {\bf B}: $B=B_{Schw}$}
\end{figure}
\clearpage

\begin{figure}
\epsscale{1}\plotone{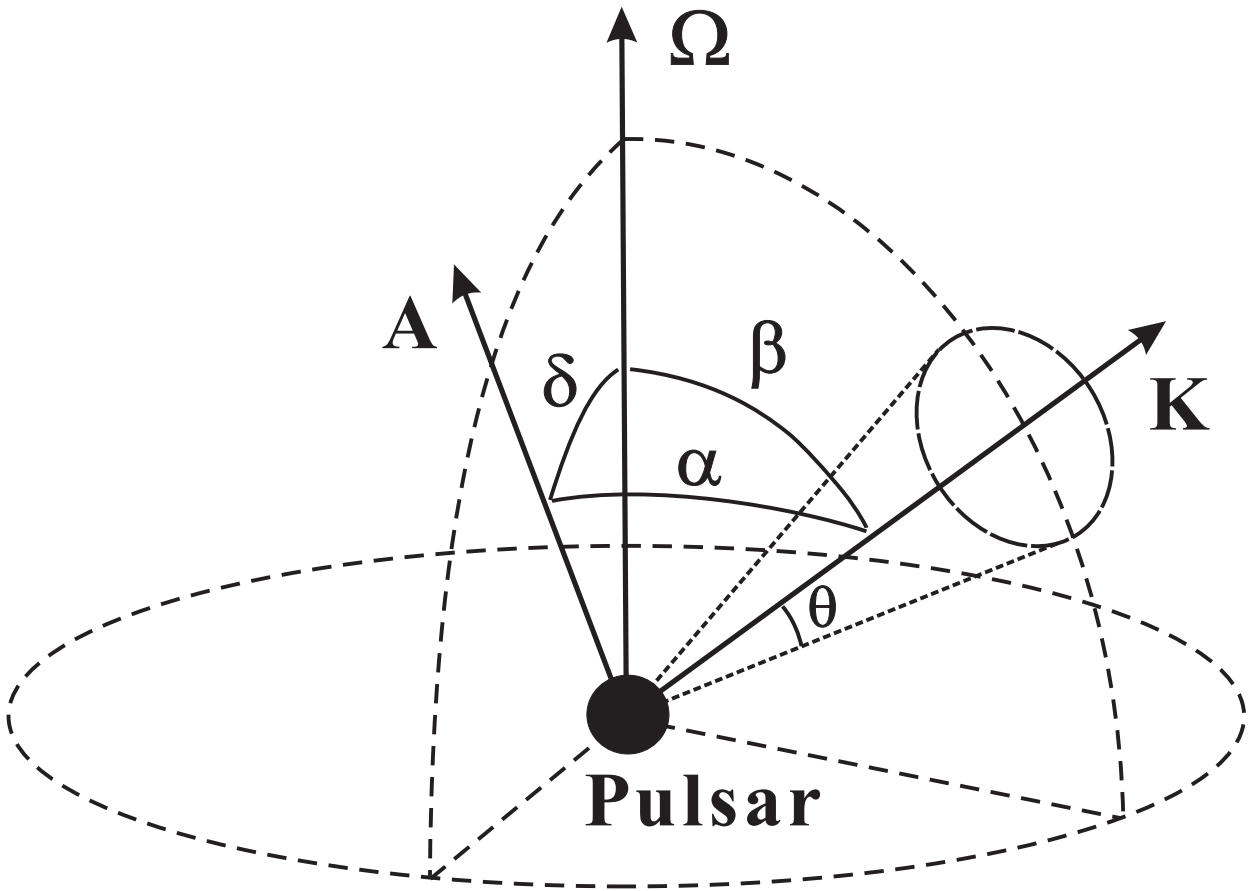} \caption{Geometry of rotation ({\bf
$\Omega$}), emission ({\bf K}), and  observers ({\bf A}) axes.
Angles $\delta$ and $\vartheta $ are constants, while $\beta$ and
$\alpha$ are oscillating with time.}
\end{figure}
\clearpage
\onecolumn
\begin{figure}
\epsscale{1}\plotone{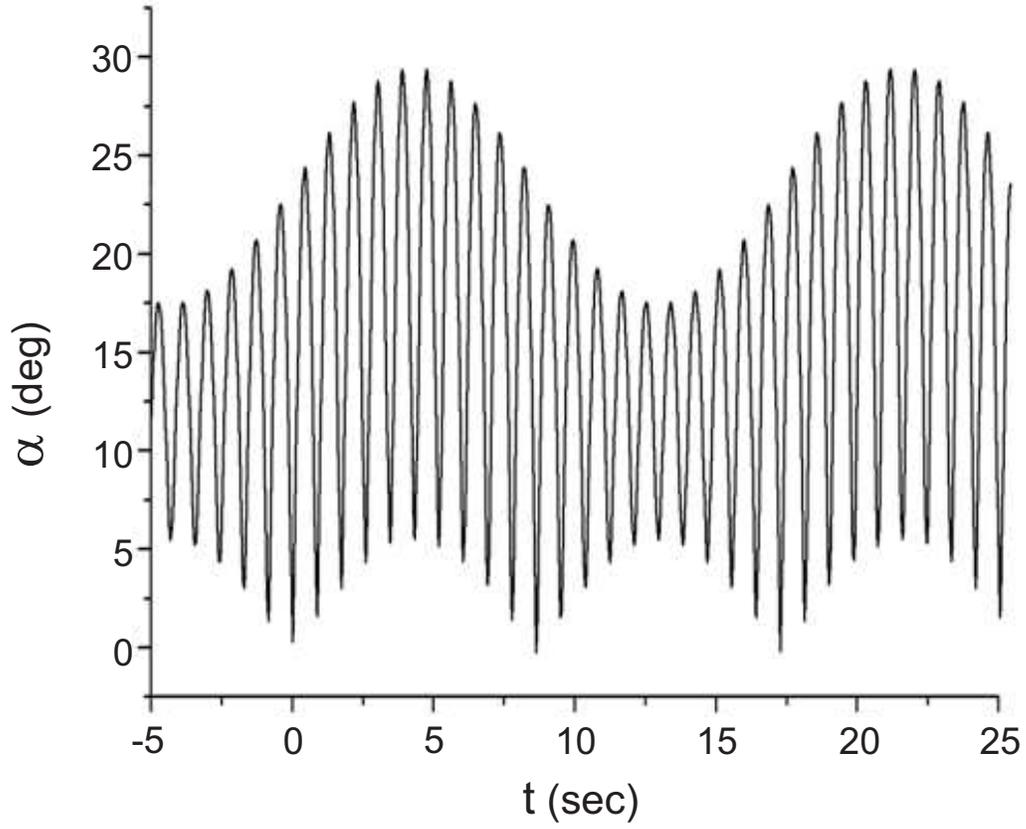} \caption{Oscillating behavior of $%
\alpha $ with time for $\beta _{0}=\delta \approx 0.12$, $\Delta
\beta=0.12$, $\omega _{dr}=2\pi/17(s^{-1})$,
$\Omega=2\pi/0.85(s^{-1})$, $\varphi=0$; $t_{1}=2\pi/\Omega$,
$t_{2}=4\pi/\Omega$}
\end{figure}

\clearpage
\begin{figure}
\epsscale{1} \plotone{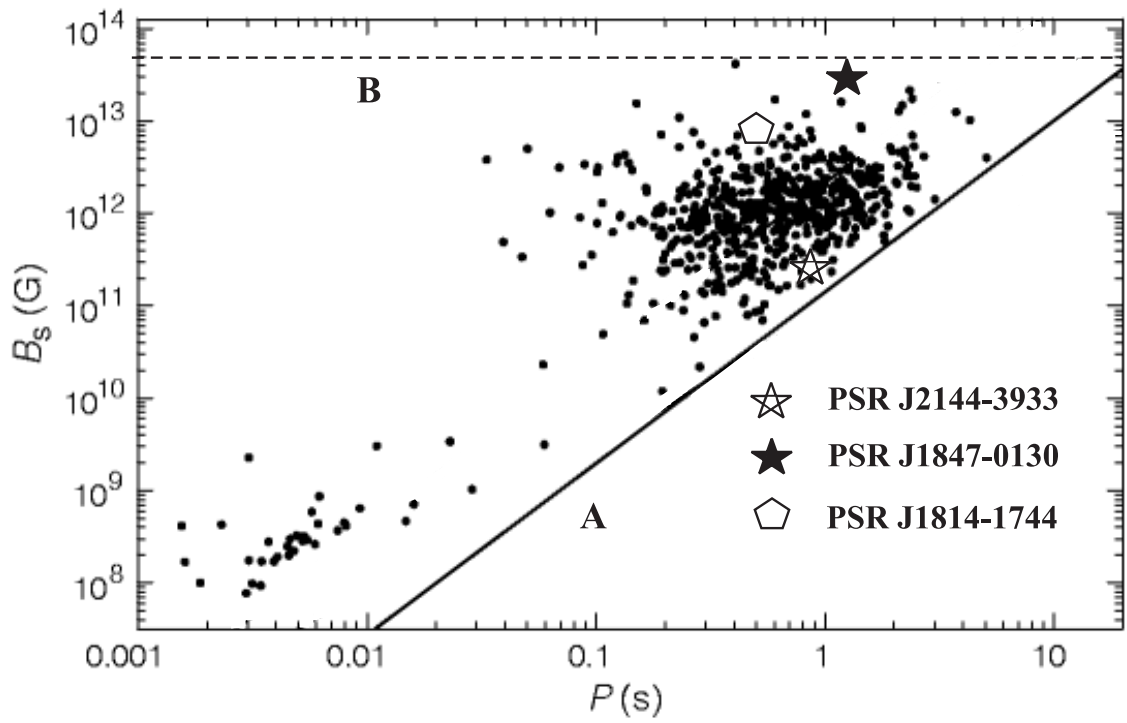} \caption{$(P-B_{s})$ diagram of the
real positions of the considered pulsars, corresponding to Table
2.}
\end{figure}

\clearpage
\begin{table}
\begin{center}
\caption{Radio pulsars with long periods.\label{tbl-1}}

\begin{tabular}{crrrrr}
\tableline\tableline No & Pulsar & $P(s)$ &
$\dot{P}(10^{-15}s\cdot s^{-1})$ & $B_{s}(10^{12}G)$ &
$\dot{E}(10^{32}erg\cdot s^{-1})$
\\
\tableline
I &PSR J2144-3933 &8.5 &0.48  &2 &0.00032 \\
II &PSR J1847-0130 &6.7 &1275 &94 &1.7  \\
III &PSR J1814-1744 &4.0 &743 &55 &4.7 \\
\tableline
\end{tabular}
\end{center}
\end{table}

\begin{deluxetable}{lcccccccccccc}
\tabletypesize{\scriptsize}
  \tablecaption{The 'real' values of pulsar parameters.\label{tbl-2}}

\tablehead{\colhead{Pulsar} & \colhead{$m$}
    & \colhead{$P_{dr}$} & \colhead{$P$}& \colhead{$\dot{P}$} & \colhead{$B_{s}$} &\colhead{$\dot{E}$}
     & \colhead{$\Delta \beta$} &\colhead{$\beta _{0} \approx \delta$} & \colhead{$\vartheta$}
     &\colhead{$W_{10}/P$}\\\colhead{}& \colhead{}
    & \colhead{(s)} & \colhead{(s)}& \colhead{$(10^{-15}s\cdot s^{-1})$} & \colhead{$(10^{12}G)$} &\colhead{$(10^{32}erg\cdot s^{-1})$}
     & \colhead{(deg)} &\colhead{(deg)} & \colhead{(deg)} &\colhead{}
    }
  \startdata
    PSR  J2144-3933 & 10 & 17.0 & 0.85 & 0.048 & 0.2 & 0.032 & $7^{o}$ & $7^{o}$ & $1.5^{o}$ & 0.1 \\
    PSR  J1847-0130 & 6 & 13.4 & 1.12 & 210 & 16 & 61 & $5^{o}$ & $5^{o}$ & $3^{o}$ & 0.3 \\
    PSR  J1814-1744 & 8 & 8.0 & 0.5 & 93 & 6.9 & 300 & $5^{o}$ & $5^{o}$ & $2^{o}$ & 0.2 \\
  \enddata
\end{deluxetable}

\end{document}